\documentclass[final,5p,times]{elsarticle}
\usepackage{graphicx}
\usepackage{amssymb}
\usepackage{amsmath}
\usepackage{multirow,array}

\usepackage{lineno,hyperref}
\usepackage{color}
\hypersetup{colorlinks=true,linkcolor=red,citecolor=cyan}
\modulolinenumbers[200]

\usepackage{multicol}

\journal{Physics Letters B}

\begin{document}

\begin{frontmatter}



\title{Particles acceleration by Bocharova-Bronnikov-Melnikov-Bekenstein black hole}


\author[1,2,3]{Bobur Turimov}
\ead{bturimov@astrin.uz}
\author[4]{Sulton Usanov}
\author[5]{Yokubjon Khamroev
}
\affiliation[1]{organization={Institute of Fundamental and Applied Research, National Research University TIIAME},
            addressline={Kori Niyoziy 39}, 
            city={Tashkent},
            postcode={100000},
            country={Uzbekistan}}
\affiliation[2]{University of Tashkent for Applied Sciences, Str. Gavhar 1, Tashkent 100149, Uzbekistan}
\affiliation[3]{Shahrisabz State Pedagogical Institute, Shahrisabz Str. 10, Shahrisabz 181301, Uzbekistan}
\affiliation[4]{Kimyo International University in Tashkent, Usman Nasyr Str.156, Tashkent 100121, Uzbekistan}
\affiliation[5]{organization={Samarkand Agroinnovations and Research University},
            addressline={Amir Temur Str. 7}, 
            city={Samarkand},
            postcode={141001},
            country={Uzbekistan}}

\begin{abstract}
{We have studied the motion of massive particles under the influence of scalar and gravitational fields, with particular emphasis on the BBMB black hole.} It has been shown that the radius of the innermost stable circular orbit (ISCO) and marginally bound orbit are significantly affected by the scalar coupling parameter. We study the energy efficiency of thin accretion disks around BBMB black holes, showing that the efficiency decreases for positive $g_s$ and increases for negative $g_s$, with a maximum of approximately $30\%$ for specific $g_s$ values. We derive analytical expressions for the angular and linear velocities of orbiting particles, highlighting their dependence on $g_s$. The photon sphere is shown to be independent of $g_s$, but the linear velocity at the ISCO position varies significantly, with massive particles behaving like ultra-relativistic particles near the black hole under scalar field influence. Additionally, we examine the center-of-mass energy (CME) of colliding particles near the BBMB black hole, showing that the scalar field can lead to infinitely high CME near the horizon, consistent with the BSW process. Astrophysical implications include CME values reaching \(10^{20} \, {\rm eV}\), comparable to the energies of ultra-high-energy cosmic rays (UHECR). 
\end{abstract}



\begin{keyword}
BBMB spacetime \sep Conformally coupled scalar field \sep BSW process \sep CME



\end{keyword}

\end{frontmatter}




\section{Introduction}

Ultra-High-Energy Cosmic Rays (UHECRs) are cosmic rays with extremely high energies, significantly exceeding those of typical cosmic rays. These particles are among the most energetic phenomena observed in the universe and are of great interest in astrophysics, particle physics, and cosmology. The production mechanisms of Ultra-High-Energy Cosmic Rays (UHECRs) with energies reaching \(10^{20} \, \text{eV}\) remain one of the most intriguing and significant challenges in modern astrophysics~\cite{Kotera2011ARAA,Zirakashivili:2023PRD}. Several mechanisms have been proposed to explain these extraordinary phenomena, including the Penrose process (PP) \cite{Penrose1971NPhS}, the Blandford-Znajek (BZ) mechanism \cite{Blandford1977MNRAS}, and the Ba\~{n}ados-Silk-West (BSW) process \cite{BSW2009PRL}. It is widely accepted that such high-energy cosmic rays are generated through the acceleration of charged particles in the presence of external magnetic fields in the vicinity of compact gravitational objects. These objects, such as Quasars, Blazars, and similar active galactic nuclei, serve as powerful astrophysical accelerators (see, e.g., \cite{Dadhich2018MNRAS,Wagh1985ApJ,Tursunov2020ApJ,Tursunov2020ApJa,Kolos2021PRD,Turimov2023PLB}). The interaction of magnetic fields with the extreme gravitational environments near these compact objects provides the necessary conditions for achieving the required particle energies. Additionally, the production of extremely high center-of-mass energies resulting from the collisions of particles near regular black holes has been investigated in detail in Refs. \cite{Patil2012PRD,Chowdhury2012PRD,Turimov2023EPJP}. These studies highlight the role of regular black holes as potential sites for high-energy cosmic ray production, further emphasizing the importance of gravitational compact objects in understanding the origins of these extraordinary astrophysical phenomena.

The motion of particles in curved spacetime is a fundamental concept in general relativity, illustrating how matter and energy influence spacetime curvature and, in turn, how this curvature governs the trajectories of objects. In such a setting, particles follow geodesics dictated by the geodesic equation. However, when a charged particle interacts with an external electromagnetic field, it undergoes acceleration and emits electromagnetic radiation. This behavior is described by a non-geodesic equation that accounts for the interaction with the electromagnetic field and the radiation reaction. Various forms of electromagnetic radiation and their mathematical representations are detailed in \cite{Landau-Lifshitz2}. Moreover, studying particles in the presence of an external scalar field is both intriguing and important. While the interaction between massive particles and scalar fields remains largely unresolved, it is theoretically crucial for understanding astrophysical processes near compact objects like black holes and neutron stars. This topic has been addressed in \cite{Noble2021NJP}, which includes an analysis of self-force effects in the presence of an external scalar field. Certain aspects of this self-force can be interpreted through spacetime geometry, shedding light on the paradox of a particle radiating without experiencing a self-force. Additionally, the acceleration of particles by black holes in the presence of a scalar field has been investigated in \cite{Zaslavskii2017IJMPD}. The interaction between scalar fields and massive particles was first introduced by Misner et al. \cite{Misner1972PRL} and later expanded upon by Breuer et al. \cite{Breuer1973PRD}, providing insights into scalar perturbations and the so-called geodesic synchrotron radiation within Schwarzschild spacetime.

In the present paper, we are interested in particle motion in the vicinity of the BBMB black hole in the presence of the scalar field. We study circular motion of test particle around the black hole and obtain characteristic radii for test particle in the presence of the scalar field. The BSW process is a theoretical mechanism in general relativity that describes how particles colliding near the event horizon of a rotating (Kerr) black hole can achieve arbitrarily high center-of-mass energies (CME) under specific conditions \cite{BSW2009PRL}. {Particles collision and extension of the BSW mechanism in several spacetime has been studied in~\cite{Feiteira:2024PRD,Tsukamoto:2024PRD,Davlataliev:2024PDU,Oteev:2024PDU,Ovcharenko:2024PRD,Turimov:2023PLB,Turimov:2023EPJP,Pryadilin:2023GC,Ovcharenko:2023PRD,Patel:2023PRD,Saha:2022JHAP,Hejda:2021PRD,Atamurotov:2021JCAP,Hackmann:2020PLB}. Using BSW process we show that the BBMB black hole might be possible high energy source. }

The paper is organized as follows: In Sec.\ref{Sec:1} we briefly describe the Einstein-scalar field-massive particle system and derive the equation of motion for the whole system. In Sec.\ref{Sec:2} we consider the circular motion of massive particles in the presence of the scalar field in the BBMB spacetime. Finally, in Sec.\ref{Sec:3} we summarize the finding results. Throughout the paper, we use geometric unit $G=c=\hbar=1$.

\section{Background geometry and particle dynamics}\label{Sec:1}

The action for the Einstein conformally coupled scalar field system can be expressed as \cite{Bocharova1970MVSFA,Bekenstein1974AP}
\begin{equation}
S=\int d^4x\sqrt{-g}\left[\frac{R}{2\kappa}-\frac{1}{2}(\nabla\Phi)^2-\frac{1}{12}R\Phi^2\right]\ ,    
\end{equation}
where $R$ is the Ricci scalar, $\Phi$ is scalar field, $g$ is the determinant of the metric tensor and $\kappa$ is Einstein constant.

The exact black hole solution in conformally coupled scalar-field gravity is described by the BBMB spacetime as follows \cite{Bocharova1970MVSFA,Bekenstein1974AP}
\begin{align}\label{metric}
ds^2=-\left(1-\frac{M}{r}\right)^2dt^2+\left(1-\frac{M}{r}\right)^{-2}dr^2+r^2d\Omega\ ,
\end{align}
where $M$ is the total mass of the black hole. The horizon of the BBMB black hole is located at $r_H=M$, which is two times less than the Schwarzschild radius. Notice that the metric \eqref{metric} is similar to the extreme Reissner-Nordstrom spacetime, however, it is associated with the scalar field given as
\begin{align}
\Phi=\sqrt{\frac{6}{\kappa}}\frac{M}{r-M}\ .    
\end{align}
which contains singularity at the horizon. 

The Lagrangian for  test particle with mass $m$ in the presence of the external scalar field $\Phi$ can be expressed as \cite{Turimov2024PRD}
\begin{align}\label{Lag}
L=\frac{1}{2}m_*g_{\mu\nu}{\dot x}^\mu {\dot x}^\nu\ ,  
\end{align}
where ${\dot x}^\mu=dx^\mu/d\tau$ is the four-velocity of the test particle normalized as ${\dot x}_\mu {\dot x}^\mu=-1$, and $\tau$ is the particle's proper time. In the above expression $m_*$ denotes the effective mass of a test particle in the presence of the external scalar field defined as
\begin{align}
m_*=m(1+\text{g}_s\Phi)\ ,
\end{align}
where $\text{g}_s$ is a coupling constant. It is important to emphasize that the effective mass of the particle also becomes divergent at the horizon like the scalar field. The four-momentum of the particle is given by 
\begin{align}\label{momentum}
p_\mu=m_*{\dot x}_\mu=m(1+g_s\Phi){\dot x}_\mu\ ,
\end{align}
Then the constants of motion, namely the specific energy ${\cal E}$ and the specific angular momentum ${\cal L}$ (energy and angular momentum of the particle per unit mass of the particle) of the test particle can be derived as
\begin{align}\label{c}
{\cal E}=(1+\text{g}_s\Phi)\left(1-\frac{M}{r}\right)^2{\dot t}\ ,\qquad {\cal L}=(1+\text{g}_s\Phi)r^2\sin^2\theta{\dot\phi}\ . 
\end{align}

\subsection{Innermost stable circular orbit}

The innermost stable circular orbit (ISCO) is an important concept in general relativity, particularly in the study of black holes and compact objects. It represents the innermost circular orbit around a massive object where stable orbital motion is possible. Beyond this orbit, circular orbits become dynamically unstable, and particles or objects are likely to spiral inward toward the central massive object. The ISCO is the radius at which the effective potential governing orbital motion has a point of inflection, meaning that small perturbations in the orbit cannot maintain stability. In astrophysical systems like black hole accretion disks, the ISCO defines the inner edge of the disk. A matter within this radius quickly plunges into the black hole. Processes near the ISCO are thought to contribute to the launching of relativistic jets in active galactic nuclei and microquasars. For binary systems involving compact objects, the ISCO marks the transition from inspiral to plunge, significantly affecting the gravitational waveforms emitted.

\begin{figure}
\centering 
\includegraphics[width=\hsize]{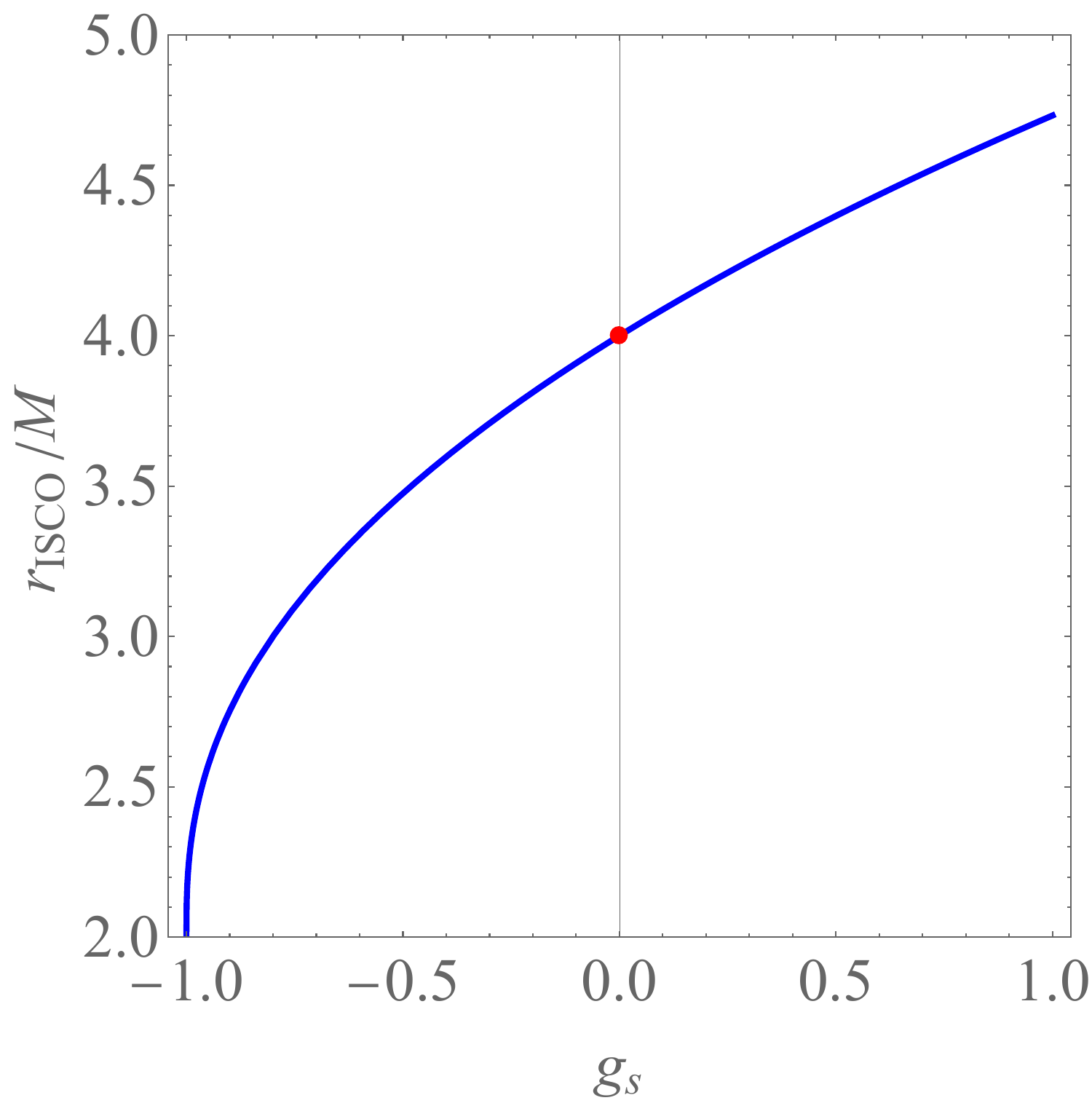}	
\caption{The ISCO radius of massive particle particle in the presence of the scalar field in the BBMB spacetime as a function of the interaction parameter $g_s$. The red dot represents the ISCO position of particle without interaction $r_{\rm ISCO}=4M$.}\label{fig_ISCO}
\end{figure}

In order to present the ISCO, we first consider motion of particle in the equatorial plane ($\theta=\pi/2$) in the vicinity of the BBMB black hole. Using normalization of the four-velocity equation for radial motion of test particle can be written as 
\begin{align}
{\dot r}^2=\frac{{\cal E}^2}{(1+g_s\Phi)^2}-\left(1-\frac{M}{r}\right)^2\left[1+\frac{{\cal L}^2}{r^2(1+g_s\Phi)^2}\right]\ .
\end{align}
Using conditions ${\dot r}={\ddot r}=0$, the critical values of conserved quantities are
\begin{align}\label{E2}
{\cal E}^2&=\frac{\left[(r-M)^2-g_sM^2\right]\left(r-M+g_sM\right)}{r^2(r-2M)}\ , \\\label{L2}
{\cal L}^2&=\frac{(1-g_s)Mr^2\left(r-M+g_sM\right)}{(r-M)(r-2M)}\ ,
\end{align}
while in absence of the interaction parameter above expressions reduce to (See e.g. \cite{Turimov2025PDU})
\begin{align}
{\cal E}^2=\frac{(r-M)^3}{r^2(r-2M)}\ , \qquad {\cal L}^2=\frac{Mr^2}{(r-2M)}\ .
\end{align}
In the ISCO, both the specific orbital energy ${\cal E}$ and the angular momentum ${\cal L}$ take their minimum values for stable circular orbits. These values are critical in accretion physics and determining the energy released during matter infall. A stationary point of the above expressions \eqref{E2} and \eqref{L2} is given by solution of a following cubic equation: 
\begin{align}\label{eqISCO}
r^3-6Mr^2+3(3-g_s)M^2r-4(1-g_s)M^3=0\ .   
\end{align}
The solution to above equation produces the ISCO radius of massive particle orbiting around the BBMB black hole in the presence of the scalar field. It is evidence that absence of the interaction parameter (i.e $g_s=0$) the ISCO position is located at $r_{\rm ISCO}=4M$ in given spacetime \cite{Turimov2025PDU}. Here we are interested in how the ISCO radius of massive particle depends on the interaction parameter. Hereafter introducing a new variable $x=r/M-2$, equation \eqref{eqISCO} reduces to 
\begin{align}
x^3-3(1+g_s)x-2(1+g_s)=0\ ,   
\end{align}
which has the following three solutions \cite{Nickalls}:
\begin{align}\label{Sol}
x_k=2\sqrt{1+g_s}\cos\left(\frac{1}{3}\cos^{-1}\frac{1}{\sqrt{1+g_s}}-\frac{2\pi k}{3}\right)\ ,     
\end{align}
for $k=0,1,2$. One can easily check that equation \eqref{Sol} is positive for $k=0$. Therefore, the ISCO position of massive particle in the presence of the scalar field is expressed as
\begin{align}\label{ISCO}
r_{\rm ISCO}=2M\left[1+\sqrt{1+g_s}\cos\left(\frac{1}{3}\cos^{-1}\frac{1}{\sqrt{1+g_s}}\right)\right]\ .    
\end{align}
From the expression \eqref{ISCO}, it is evident that the minimum value of the interaction parameter is $g_s = -1$. At this value, the ISCO position of a particle around the BBMB black hole approaches $2M$. In Fig.\ref{fig_ISCO}, we illustrate the dependence of the ISCO radius for a massive particle orbiting the BBMB black hole in the presence of a scalar field. The figure shows that the ISCO position increases for positive values of the scalar parameter, whereas it decreases for negative values of the scalar parameter.

\subsection{Marginally bound orbits}

\begin{figure}
\centering 
\includegraphics[width=\hsize]{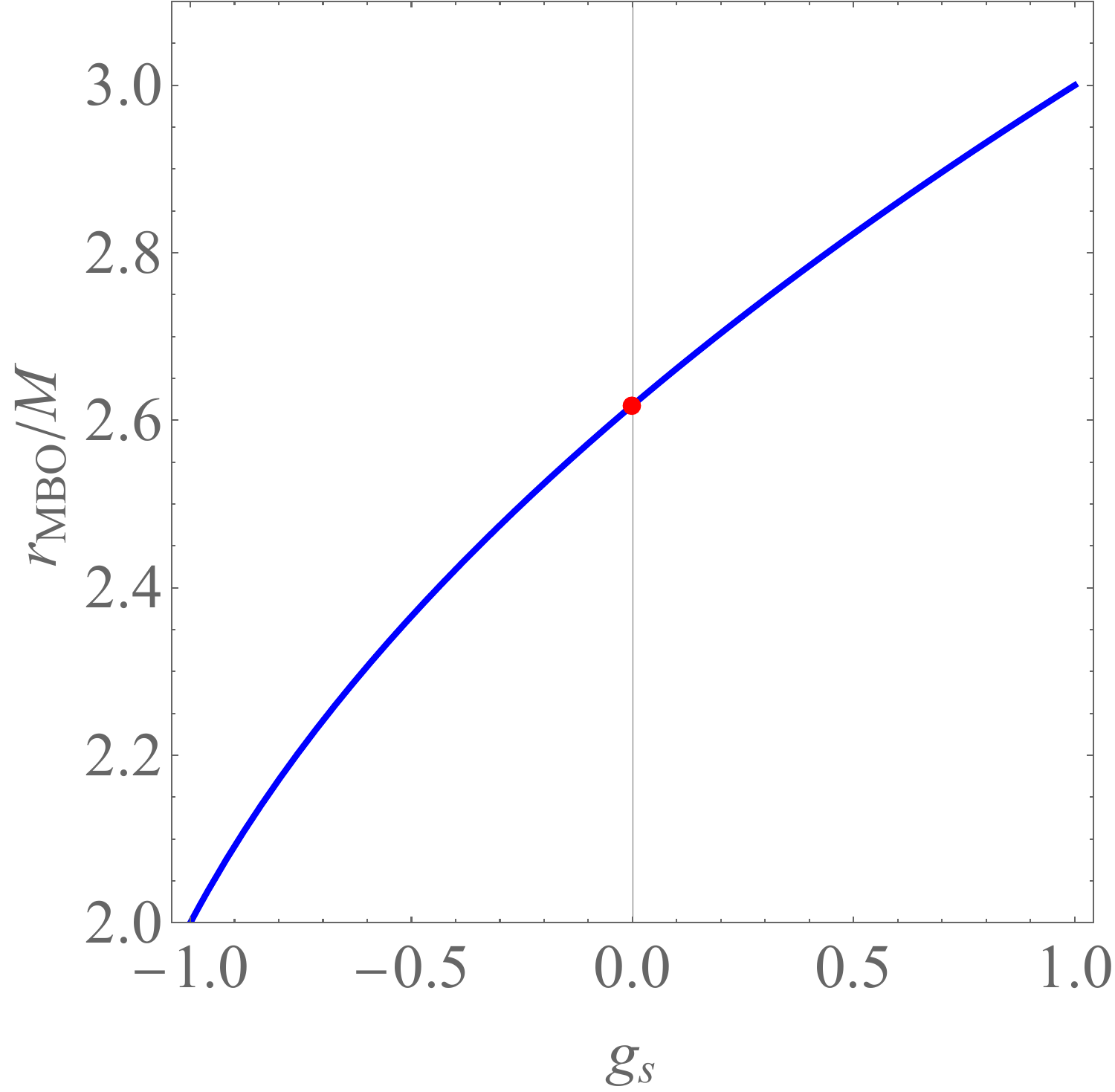}	
\caption{The MBO radius of massive particle particle in the presence of the scalar field in the BBMB spacetime as a function of the interaction parameter $g_s$. The red dot represents the MBO position of particle without interaction $r_{\rm MBO}=(3+\sqrt{5})/2$.}\label{fig_MBO}
\end{figure}

A marginally bound orbit (MBO) refers to a specific type of orbit massive particle in the gravitational field. It represents the critical boundary between bound and unbound motion, where a particle or object is just barely gravitationally bound to the central massive object. A massive particle in MBO has just enough kinetic energy to escape to infinity if slightly perturbed, but will remain gravitationally bound otherwise. The specific energy of massive particle in the MBO is exactly equal to one (i.e. ${\cal E}=1$). Now, by normalizing the specific energy in \eqref{E2}, the MBO radius for test particle can be found as follows: 
\begin{align}
r_{\rm MBO}=\frac{1}{2}\left(3+\sqrt{5+4g_s}\right)\ ,    
\end{align}
which reduces to $r_{\rm MBO}=(3+\sqrt{5})/2$ without taking into account interaction parameter (i.e. $g_s=0$). In Fig.\ref{fig_MBO}, we present the dependence of the MBO radius on the interaction parameter. It is evident that the MBO position for particle increases for positive values of the scalar parameter, whereas it decreases for negative values of the scalar parameter.

\subsection{Energy efficiency}

\begin{figure}
\centering 
\includegraphics[width=\hsize]{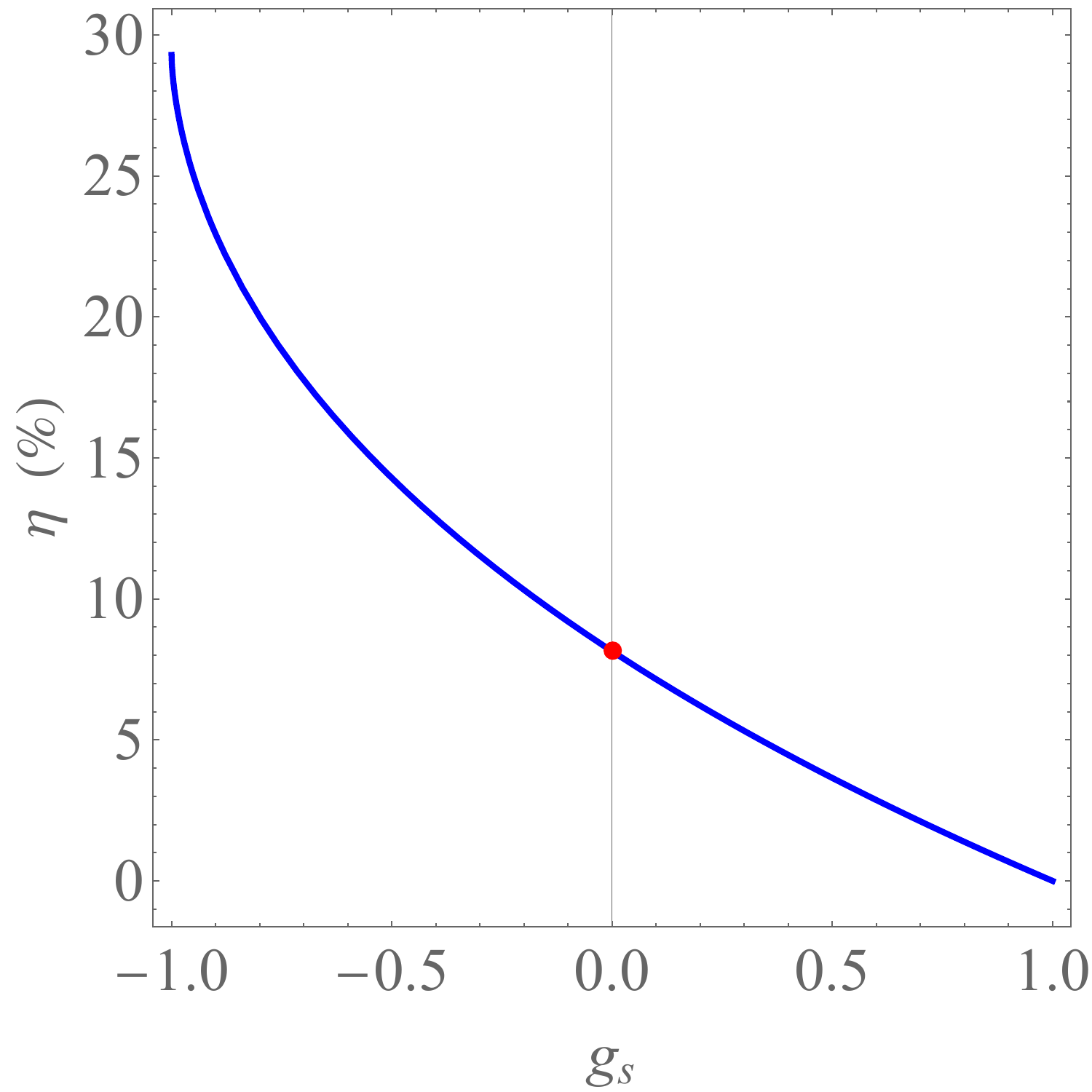}	
\caption{The energy efficiency of massive particle particle in the presence of the scalar field in the BBMB spacetime as a function of the interaction parameter $g_s$. The red dot represents the energy efficiency of particle without interaction $\eta\simeq8\%$.}\label{fig_eta}
\end{figure}

The Novikov-Thorne model provides a framework for describing a thin accretion disk around black holes, including the energy efficiency of the accretion process. This model is often used in astrophysics to study thin, relativistic accretion disks in general relativity. The energy efficiency of a black hole accretion process is defined as the fraction of the rest mass energy of the accreted matter that is converted into radiation and escapes to infinity: $\eta=1-{\cal E}_{\rm ISCO}$, where $\eta$ is the efficiency of energy conversion and ${\cal E}_{\rm ISCO}$ is the specific energy of matter at the ISCO. In Fig.\ref{fig_eta}, we illustrate the dependence of the energy efficiency of a particle in the BBMB spacetime influenced by the scalar field. The energy efficiency noticeably decreases for positive values of the scalar parameter and increases for negative values. In particular, the maximum energy efficiency reaches to approximately $30\%$ at the minimum value of the interaction parameter (i.e. $g_s=-1$).

\subsection{Angular and orbital velocity}

It is interesting to consider the angular velocity of particle $\Omega$ orbiting around the BBMB black hole in the presence of the scalar field. Now we focus on how the angular velocity of depends on the geometry of the BBMB spacetime and the interaction parameter. Using the following expression
\begin{align}
\Omega=\frac{d\phi}{dt}=\frac{1}{r^2}\left(1-\frac{M}{r}\right)^2\frac{\cal L}{\cal E}\ ,
\end{align}
the angular momentum of test particle orbiting around the BBMB black hole is expressed as
\begin{align}\label{Omega}
\Omega=\sqrt{\frac{M}{r^3}\left(1-\frac{M}{r}\right)\frac{1-g_s}{1-g_s\left(\frac{r}{M}-1\right)^{-2}}}\ .    
\end{align}
In the case when $g_s=0$, the expression \eqref{Omega} reduces to the following simple form $\Omega=\sqrt{M/r^3(1-M/r)}$ which represents the angular velocity of test particle orbiting around the BBMB black hole \cite{Turimov2025PDU}.

\begin{figure}
\centering 
\includegraphics[width=\hsize]{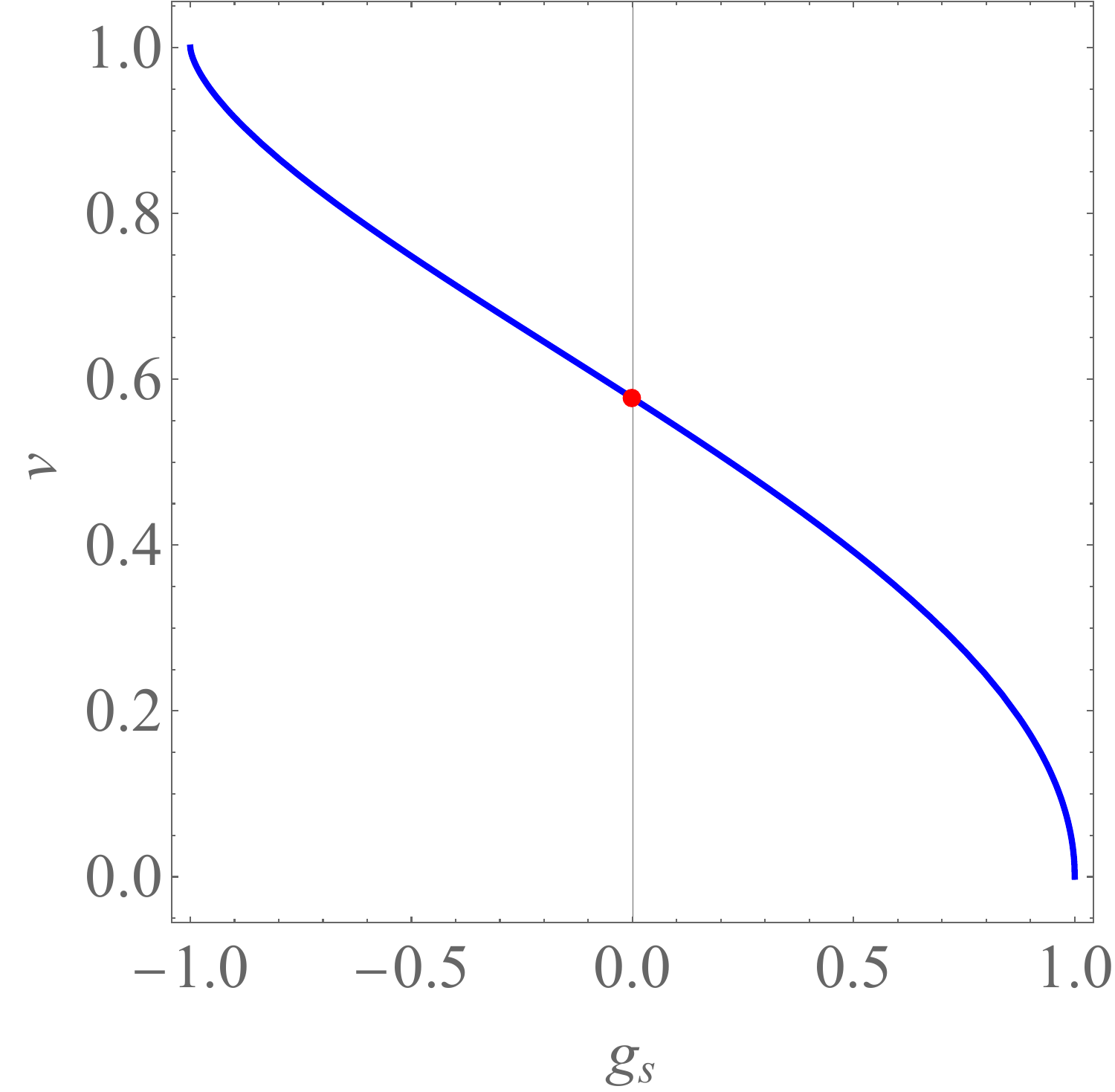}	
\caption{The orbital velocity of massive particle particle orbiting around the BBMB in the presence of the scalar field as a function of the interaction parameter $g_s$. The red dot represents the orbital velocity of particle without interaction $v=1/\sqrt{3}$.}\label{fig_vel}
\end{figure}

It is also interesting to consider linear velocity of massive particle orbiting the BBMB black hole. Using the following definition, $v=\Omega\sqrt{-g_{\phi\phi}/g_{tt}}$ (See e.g., \cite{Turimov2022Universe}), the linear velocity of massive particle in the BBMB spacetime in the presence of the scalar is determined as
\begin{align}\label{v}
v=\sqrt{\frac{1-g_s}{\frac{r}{M}-1-g_s\left(\frac{r}{M}-1\right)^{-1}}} \ .   
\end{align}
From equation \eqref{v}, it is evident that the orbital velocity of a test particle becomes equal to the speed of light (or one in natural units) at $r = 2M$. In our earlier work \cite{Turimov2025PDU}, we demonstrated that $r = 2M$ corresponds to the location of the photon sphere. Notice that the motion of photons is completely independent of the interaction parameter $g_s$. Figure \ref{fig_vel} illustrates the dependence of the linear velocity of a test particle orbiting at the ISCO position around the BBMB black hole on the interaction parameter $g_s$. In previous analyses, we established that the minimum value of the interaction parameter is $g_s=-1$, at which the orbital velocity reaches the speed of light. According to this fact, one can conclude that massive test particle behaves like ultra-relativistic particle near the BBMB black hole in the presence of the scalar field.

\section{BSW mechanism}\label{Sec:2}

According to Ref. \cite{BSW2009PRL}, the center-of-mass energy (CME) for the colliding pair of particles in curved spacetime
\begin{align}\label{E}
E^2_{cm}&=-\left(p_{1\mu}+p_{2\mu}\right)\left(p_1^{\mu}+p_2^{\mu}\right)\nonumber\\&=-p_{1\mu}p_1^{\mu}-p_{2\mu}p_2^{\mu}-2p_{1\mu}p_2^{\mu}\ .
\end{align}
Using the definition of the four-momentum in \eqref{momentum}, one can obtain $p_{i\mu}p_i^\mu=-m_{*i}^2=-m_{i}^2(1+g_{si}\Phi)^2$, with $i=1,2$. The last term of the equation \eqref{E} can be simplified as
\begin{align}\nonumber
p_{1\mu}p_2^{\mu}&=m_1m_2(g^{tt}{\cal E}_1{\cal E}_2+g^{\phi\phi}{\cal L}_1{\cal L}_2)\\&+m_{*1}m_{*2}(g_{rr}{\dot r}_1{\dot r}_2+g_{\theta\theta}{\dot\theta}_1 {\dot\theta}_2)\ .   
\end{align}
Finally, the center-of-mass energy of colliding particles in the equatorial plane ($\theta=\pi/2$) of the black hole can be determined as
\begin{align}\label{Ecm}\nonumber
E^2_{cm}&=m_1^2(1+g_{s1}\Phi)^2+m_2^2(1+g_{s2}\Phi)^2\\&+2m_1m_2\left[\frac{{\cal E}_1{\cal E}_2-Y_1Y_2}{(1-M/r)^2}-\frac{{\cal L}_1{\cal L}_2}{r^2}\right]\ ,
\end{align}
where 
$$
Y_i=\sqrt{{\cal E}^2_i-\left(1-\frac{M}{r}\right)^2\left[(1+g_{si}\Phi)^2+\frac{{\cal L}_i^2}{r^2}\right]}\ .
$$
As one can see from equation \eqref{Ecm} that the CME of colliding particles divergent at the horizon. To avoid this issue given equation should be modified and after simple algebraic manipulations, equation \eqref{Ecm} can be rewritten as
\begin{align}\nonumber
&\frac{E^2_{cm}}{m_1m_2}=\frac{m_1}{m_2}(1+g_{s1}\Phi)^2+\frac{m_2}{m_1}(1+g_{s2}\Phi)^2-\frac{2{\cal L}_1{\cal L}_2}{r^2}\\\nonumber&+\frac{2}{{\cal E}_1{\cal E}_2+Y_1Y_2}\left\{\left(1-\frac{M}{r}\right)^2\left[\frac{{\cal L}_1^2}{r^2}+(1+g_{s1}\Phi)^2\right]\left[\frac{{\cal L}_2^2}{r^2}+(1+g_{s2}\Phi)^2\right]\right.\\&+\left.{\cal E}_2^2\left[\frac{{\cal L}_1^2}{r^2}+(1+g_{s1}\Phi)^2\right]+{\cal E}_1^2\left[\frac{{\cal L}_2^2}{r^2}+(1+g_{s2}\Phi)^2\right]\right\}\ ,
\end{align}
and near the horizon of the BBMB black hole

\begin{align}\nonumber
&\frac{E^2_{cm}(r\to r_H)}{m_1m_2}=\frac{1}{M^2}\left({\cal L}_2\sqrt{\frac{{\cal E}_1}{{\cal E}_2}}-{\cal L}_1\sqrt{\frac{{\cal E}_2}{{\cal E}_1}}\right)^2\\&+\left(\frac{m_1}{m_2}+\frac{{\cal E}_2}{{\cal E}_1}\right)(1+g_{s1}\Phi_H)^2+\left(\frac{m_2}{m_1}+\frac{{\cal E}_1}{{\cal E}_2}\right)(1+g_{s2}\Phi_H)^2\ ,
\end{align}
where a sub-index $H$ denotes the values of the quantities at the horizon. In the case of identical particles with $m=m_1=m_2$ and ${\cal E}_1={\cal E}_2$, one gets
\begin{align}\label{ECM2}
&\frac{E^2_{cm}(r\to r_H)}{m^2}=\frac{\left({\cal L}_2-{\cal L}_1\right)^2}{M^2}+4(1+g_{s}\Phi_H)^2\ .
\end{align}
It is evident that the first term of the expression \eqref{ECM2} is finite, while the second term tends to infinity due to divergence of the scalar field at the horizon. It turns out that the CME of colliding particles produce infinity energy with arbitrary unit.

\subsection{Application on ultra-high-energy cosmic ray}

\begin{figure}
\centering 
\includegraphics[width=\hsize]{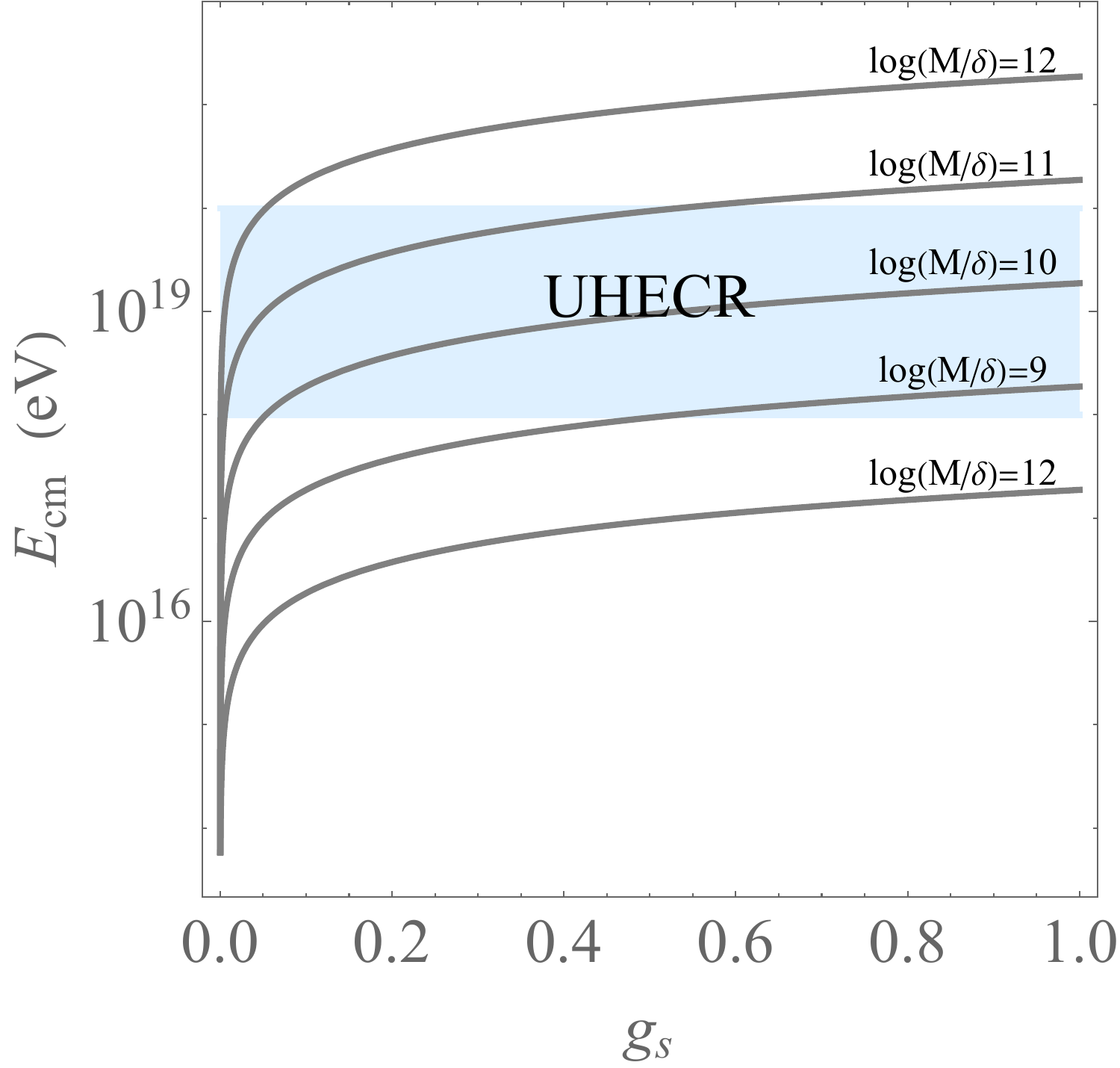}	
\caption{The CME of binding particles in the vicinity of the BBMB black hole as a function of interaction parameter $g_s$ for the different positions from the black hole horizon. A shaved region represents observational evidence of UHECR with energy of $\sim 10^{18}-10^{20} \, {\rm eV}$.} \label{fig_CME}
\end{figure}

The first detection of a cosmic-ray particle with an energy exceeding $1.0 \times 10^{20} \, {\rm eV}$ in the Volcano Ranch experiment \cite{Linsley1963PRL}. One has to emphasize that such huge energy can be obtained for the CME of colliding particles in the vicinity of the BBMB black hole. The first term of equation \eqref{ECM2} is finite, therefore it can be safely neglected. Let us now consider the CME of particles around the horizon of the BBMB black hole: $r=r_h+\delta$, where $\delta$ is the small distance from the black hole horizon. The CME of colliding particles in the vicinity of the BBMB black hole reads  
\begin{align}\label{E}
E_{cm}(\delta)\simeq 2m\left(1+g_s\frac{M}{\delta}\right) \ .   
\end{align}
The dependence of the CME energy of particles from the interaction parameter $g_s$ for the different values of the mass-distant ratio $M/\delta$ is illustrated in Fig.\ref{fig_CME}. It is evidence that the distance $\delta$ can be determined as 
\begin{align}
\delta\simeq 150\left(\frac{M}{10^9M_\odot}\right)\,{\rm m}\ ,    
\end{align}
for energy of $10^{19} {\rm eV}$.

\section{Conclusions}\label{Sec:3}

In this paper, we have examined the motion of a massive particle influenced by scalar and gravitational fields, with particular attention to the BBMB spacetime. By deriving the equations of motion and the effective potential for a massive particle, we investigated the ISCO radius in this spacetime. The ISCO radius, a key factor in understanding the dynamics around compact objects, was found to be affected by the scalar field through the interaction parameter $g_s$. Our results reveal that the ISCO radius increases for positive $g_s$ and decreases for negative $g_s$, underscoring the substantial role of scalar interactions in shaping orbital dynamics. A very similar result has been obtained for the MBO of test particle in the vicinity of the BBMB black hole in the presence of the scalar field. 

It is studied energy efficiency of the BBMB black hole using the Novikov-Thorne model for studying thin accretion disks around black holes, particularly in relativistic contexts. In the BBMB spacetime, influenced by a scalar field, the energy efficiency is shown to vary with the scalar parameter $g_s$, decreasing for positive values and increasing for negative values. Notably, the maximum energy efficiency approaches approximately $30\%$ for particular value of interaction parameter $g_s$, highlighting the significant impact of scalar fields on accretion dynamics.

The angular and linear velocities of particles orbiting the BBMB black hole in the presence of a scalar field provide valuable insights into the dynamics of spacetime geometry and the scalar field interaction has been studied. The analytical expressions for both quantities are presented. It has been shown that the photonsphere is independent of the interaction parameter. The dependence of the linear velocity at the ISCO position on $g_s$ investigated. It is also shown that massive test particles can behave like ultra-relativistic particles near the BBMB black hole under the influence of a scalar field.

{We examined the CME of binding massive particles in the vicinity of the BBMB black hole, influenced by the presence of a scalar field. According to the BSW process, a static black hole can produce a high-energy particle \cite{BSW2009PRL}. In the present paper, we have shown that in the presence scalar field the CME of colliding particles near the BBMB black hole can tends to infinity. We derived the exact expression for the CME of colliding particles in the presence of the scalar field. We found that the presence of a scalar field significantly alters the CME of particles, particularly near the horizon of the BBMB black hole. As an astrophysical consequence we have found that the CME of colliding particles reach $10^{20}$ in the vicinity of the black hole which is an observable energy of UHECR. Our findings provide insights into the complex behavior of particles near compact objects, emphasizing the importance of considering scalar fields in relativistic astrophysics. This work contributes to a deeper understanding of the high-energetic sources and potential observational signatures of particles in scalar-field-influenced spacetimes. In future we are planing to consider BSW mechanism in the rotating black hole spacetime in conformally coupled scalar field theory.}


\appendix

\bibliographystyle{model}  
\bibliography{example}





\end{document}